# Lattice Fusion Networks for Image Denoising

Seyed Mohsen Hosseini, smhosseini741@gmail.com

## Abstract

*A novel method for feature fusion in convolutional neural networks is proposed in this paper. Different feature fusion techniques are suggested to facilitate the flow of information and improve the training of deep neural networks. Some of these techniques as well as the proposed network can be considered a type of Directed Acyclic Graph (DAG) Network, where a layer can receive inputs from other layers and have outputs to other layers. In the proposed general framework of Lattice Fusion Network (LFNet), feature maps of each convolutional layer are passed to other layers based on a lattice graph structure, where nodes are convolutional layers. To evaluate the performance of the proposed architecture, different designs based on the general framework of LFNet are implemented for the task of image denoising. This task is used as an example where training deep convolutional networks is needed. Results are compared with state of the art methods. The proposed network is able to achieve better results with far fewer learnable parameters, which shows the effectiveness of LFNets for training of deep neural networks.*

## 1. Introduction

Deep neural networks have been very effective in different machine learning tasks. Superior results in image classification challenges as well as other image recognition problems are based on deep network structures [1, 2, 3]. The depth of the deep network plays an important role in its success [1, 4]. Deeper networks with more layers are able to extract and integrate more levels of hierarchical features. But deeper networks are more difficult to train. In deep networks the input data has to pass a large number of layers to reach the output layer, and in the opposite direction the gradient reaches the beginning of the network after passing many layers. This will cause the vanishing or exploding gradient problem. When the gradient is very small in the initial layers the training will not be very effective. A lot of difference designs like, ResNet [5], DenseNet [6], Highway Network [7], and FractalNet [8], have been proposed to improve the training of deep networks. These methods try to facilitate the flow of gradient to initial layers and flow of input information to deeper layers, through skip connections and feature fusion between different layers at different depths. These networks can be considered special cases of Directed Acyclic Graph (DAG) Network where a layer can receive input from multiple layers and pass its feature maps to multiple layers. Based on the different strategies to connect different layers in a DAG Network different models are formed. In the proposed network layers pass their feature maps based on a lattice graph structure where nodes are convolutional layers. Figure 1 shows a Lattice Fusion Network (LFNet) which 20 layers arranged in a lattice graph structure with 4 rows and 5 columns. With this strategy there are multiple possible paths from input to the output layer, so there are different possible depths for the information to pass through. To evaluate the performance of the proposed network different designs based on the general framework of LFNet is implemented for the task of image denoising. The proposed method is tested using several commonly used gray and color image datasets. Better results compared to state of the art methods are achieve with far fewer learnable parameters which shows the effectiveness of the proposed framework for training of deep networks. LFNet is able to outperform the well known stat of the art DnCNN [12] with half (52%) the number of learnable parameters.

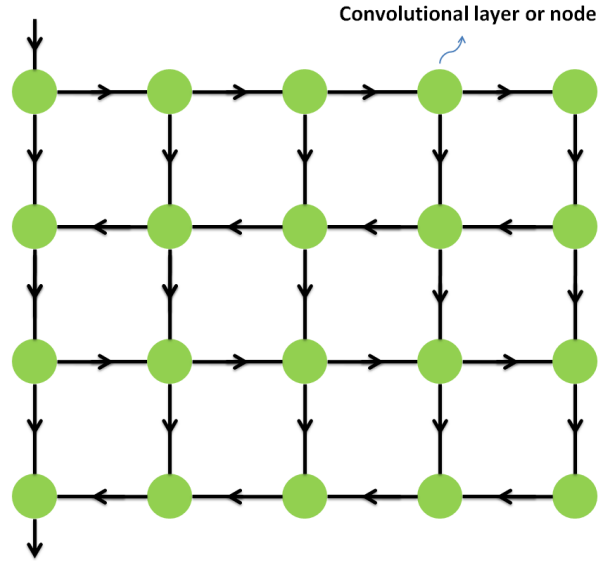

Figure 1. An example of a Lattice Fusion Network (LFNet) with 20 convolutional layers or nodes arranged in a lattice graph structure with 4 rows and 5 columns.

## 2. Related work

Deep neural networks have produced superior results in high-level computer vision tasks such as image recognition. They also have been very effective in low-level image denoising and restoration tasks. Because of the difficulties of training deeper networks, like vanishing or exploding gradient, simply stacking more layers does not necessarily lead to better results and at some point the performance degrades [5]. Different techniques have been proposed to improve the training of deep networks. Skip connection is a very common technique to ease the flow of gradient to initial layers. It also facilitates the flow of information from input and information produced by initial layers to deeper layers. Skip connection is used in ResNet [5], DenseNet [6], Highway network [7], and in U-Nets [13] in different ways.

The ResNet [5] structure consists of a number of residual learning building blocks. Each block consists of two convolutional layers and two ReLU layers, with input being added to the output of the second convolutional layer with an identity connection. This model is a combination of residual learning and shortcut connection which have been used in other contexts [9, 10]. In ResNet the layers are trained based on the residual values and this has enabled successful training of very deep networks which have produced state of the art results in many image recognition and detection challenges. The depth can be increase further by stochastic depth method [11] which is randomly removing layers during the training. With this method ResNets with more than 1200 layers can be trained and produce improved results.

Highway network [7] employs skip connections with gating mechanism. The gating mechanism has learnable parameters and responds to different input data differently, unlike a skip connection, used in ResNet, that never closes. In DenseNet [6] the number of connections between layers is further increased compared to ResNet. DenseNet consists of dense blocks. In a dense block the output of a layer is concatenated with the input of the next layer, so the last layer in each block receives the feature maps of all the layers in the block. This structure enables creating features from low and high level feature maps. DenseNet achieves 5.24% error on CIFAR-10 with depth of 40 and 1.0M parameters compared to the 6.61% error of ResNet with depth of 110 and 1.7M parameters. The error of DenseNet can be further reduced to 3.46% with DenseNet-BC with depth of 190, but with 25.6M parameters which is more than 15 times of the number of parameters of ResNet. So at certain depths the DenseNet is more efficient but in deeper networks the improvement in performance comes at the expense of a larger number of parameters.

FractalNets [8] also consist of a number of building blocks. In each block the output of multiple parallel paths with different number of layers are joined repeatedly. This structure creates multiple possible paths for information to go through. The join layer merges a number of feature maps into one. Element-wise mean is chosen as the join layer, and they mention concatenation and addition as other possible choices for the join layer. The error of FractalNet on CIFAR-10 is 5.22% with the depth of 21 and 38.6M learnable parameters, which is an improvement of 0.02% compared to DenseNet with depth of 40, but with 38.6 times the parameters.

In the task of image denoising or image restoration the goal is to recover a clean image from a noisy or degraded one. These tasks include removing Additive White Gaussian Noise (AWGN), single image super resolution, and JPEG image deblocking. Methods based on convolutional networks with residual learning have been very successful in this task producing state of the art results. In these methods the network is trained on pairs of noisy images and the noise. The trained network would map a noisy input to noise; this noise then is removed from the input producing a denoised or recovered image.

In [12] a denoising convolutional neural network based on residual learning and batch normalization [15] called DnCNN is proposed. Batch normalization reduces the internal covariate shift and improves the speed of convergence as well as performance. In Gaussian denoising the output image (Gaussian noise) and batch normalization both are related to Gaussian distribution and it is shown in [12] that the combination of batch normalization and residual learning improves the results. DnCNN consists of a number of convolutional layers with batch normalization and ReLU layers stacked one after the other. The design of DnCCN is relatively simple but the network is very effective in different denoising tasks. DnCNN-S is a non-blind denoising network (different networks are used for specific noise levels). This network consists of a stack of 17 convolutional layers. DnCNN-B is a blind denoising network (a single network is used for a range of noise levels). This network has 20 convolutional layers, and can be used for noise levels in the range of [0,55]. Number of filters in DnCNNs is 64 except for the last layer which has one filter for gray image denoising and three for color image denoising.

To have a more meaningful comparison residual learning and batch normalization are employed in the LFNet architectures implemented here for the image denoising.

Mapping the noisy image to noise requires the network to learn to separate noise from the latent clean image in the input. Learning the latent image is easier with more hierarchical features and thus more layers, so deep networks are popular in image denoising tasks.

Beside more hierarchical features having larger receptive field is another advantage of deep networks. With larger receptive field more contextual information is extracted from the image and learning the latent clean image becomes easier. So, to achieve higher performances deeper networks should be used. But because of the difficulties of training a deep network, the performance does not increase by simply stacking more layers in a plain network. The vanishing gradient problem would be more severe in deeper plain networks. But apart from that, if the information of the initial layers is not passed to deeper layers, image details could be lost because the output is reconstructed with image abstractions, this leads to performance degradation [14]. To overcome these training difficulties, LFNet architectures are used here to have a deep network that can be trained effectively and produce superior results.

## 3. Lattice Fusion Networks (LFNets)

The main idea of the proposed network is to have a short distance between each layer and output while at the same time having a long depth for feature maps to go through. In a directed acyclic lattice graph with $n$ rows and $m$ columns ($n < m$) the shortest distance between input and output is $n$ ($depth = n$) at the same time the data can go through all of the nodes before reaching the output layer ($depth = n \times m$). So the number of layers that the input can pass through ranges from $n$ to $n \times m$. In LFNets convolutional layers are connected based on a lattice graph structure. Each convolutional layer or node receives maximum of 2 sets of feature maps from its neighbors and outputs its feature maps to maximum of 2 of its neighboring convolutional nodes. Different sets of feature maps received by a node are concatenated to form its input. Other combining methods like adding can also be implemented to keep the number of learnable parameters down. What follows is a list of main characteristics of LFNets and their differences with other feature fusion methods.

- Multiple depths are possible in a single LFNet. Figure 2 shows 3 routs as an example of different possible routes that the information can propagate through in a single network, depths of 4 to 16 are present in this network. As opposed to other methods, in LFNet the different possible paths of information propagation are not manually designed, but they are a feature of a lattice graph structure.

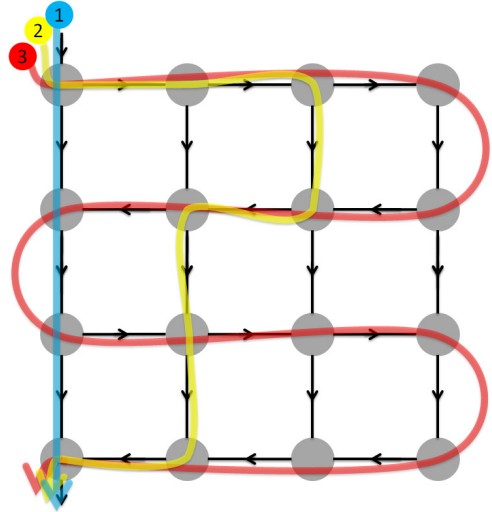

Figure 2. Feature maps created by different layers have multiple possible routes to reach the output layer in a Lattice Fusion Network (LFNet). Three routes are shown here as an example. There are 16, 8 and 4 layers in routs 3, 2 and 1 respectively. In this specific network 4 is the minimum depth and 16 is the longest maximum depth.

- In LFNet the distance between layers and output is much less than the maximum possible depth of the network. This will facilitate more efficient gradient propagation. In the network of Figure 2 the maximum depth is 16 but the maximum distance to output is 6 layers, which is for the layer at the upper right corner of the network. Other layers have a distance of less than 6 layers to output.

- In LFNet because of the way that feature maps are passed to other layers there is a strong interaction between features of different levels, and the output can be produced using low, mid, and high level feature maps. Some existing feature fusion methods are based on a building block structure. Feature fusion mostly happens between layers inside of a block and the result is then passed to the next block. So the interactions are mostly limited to individual blocks and the design of the network.

- There is no need for block structure in LFNet. In some existing methods in order to keep the number of learnable parameters down the network should be divided to a number of blocks. Even with block structure the number of parameters can increase rapidly. For example, in DenseNet the layers at the end of a block receive multiple inputs from all of the previous layers of the block. With this design having more layers in a block can lead to rapid increase of parameters. In LFNet there is no possibility for rapid increase of parameters. Each layer receives and outputs feature maps to maximum 2 of its immediate neighbors.

- The building block structure of existing methods creates a bottle neck effect for the flow of information. All the feature maps inside of a block must be reduced and combined to form the output of the block. In LFNet the connection pattern of the layers is uniform throughout the network and there is no bottleneck for the propagating of information.

- In LFNet different parts of the structure can have different convolutional layers, for instance, layers with different filter sizes or dilation factors. Different layers can complement each other without interrupting each other's path to input and output layers.

- The architected of LFNet is easily scalable to networks with more layers. In the existing popular methods the distance of layers to the output is dependent on the number of blocks in the network, so in a deeper network with more blocks, layers are farther away from output which makes the propagation of gradient more difficult. But in LFNet the distance of the layers to the output is more dependent on their position in the graph and not the number of other layers. For example in the network of Figure 2 the farthest layer from output is the layer at upper right corner and its distance to output is 6 layers. In a network with 12 more layers (3 more columns) the distance of the farthest layer from output (the layer at upper right corner) would be 9 layers. So, the second network has 12 more layers but the maximum distance to output is only increased by 3 layers. This makes LFNet an effective structure for very deep networks.

## 4. Results

To evaluate the performance of the proposed network in training of deep networks, a few networks based on LFNet architecture were trained and tested for the task of image Gaussian denoising. For gray image denoising an LFNet with 4 rows and 5 columns (LFNet_4-5) and an LFNet with 4 rows and 6 columns (LFNet_4-6) were trained.

These networks also have an output layer, so the total number of layers is 21 and 25 respectively. In LFN_4-6 the minimum depth in the network is 5 and the maximum depth is 25 (total number of layers in the network). The training conditions of the networks were chosen similar to the training conditions of DnCNN [12] to have a more meaningful comparison. Training images were 400 images of size $180 \times 180$ cropped from BSDS500 data set (train and test images) [21]. Gaussian noise is added to the image patches to form noisy patches which are the input of the network. The residual patches (noise itself) are used as target images (residual learning). Each node in the LFN contains a convolution layer, a Batch normalization layer and a ReLU activation layer. Zero padding is utilized to keep the feature size the same. In the nodes that receive input from two other nodes concatenation is used. All of the convolutional layers have 32 filters. The input layer (first layer) has 32 filters of size $3 \times 3 \times 1$. Convolutional layers in the nodes that receive one input have 32 filters with size $3 \times 3 \times 32$. Convolutional layers in the nodes that receive two sets of feature maps from neighboring nodes have 32 filters with size $3 \times 3 \times 64$. The output layer (last layer) has 1 filter of size $3 \times 3 \times 64$. The receptive field of a network with depth d and filter size $3 \times 3$ is $(2d+1) \times (2d+1)$. The total depths of LFN_4-5 and LFN_4-6 are 21 and 25 respectively, and given the size of the filters the receptive fields are 43 and 51 respectively. The patch sizes are chosen $50 \times 50$ for FLN_4-5 and $60 \times 60$ for LFN_4-6. Larger patch sizes bring limited improvement of performance with higher computational cost. In [12] the patch size for the 20 layer DnCNN-B is set to $50 \times 50$. $128 \times 1600$ pairs of noisy and residual patches were used as the training data. In gray image denoising different networks were trained for specific noise levels. Adam optimization algorithm was used and the learning rate was reduced from 1e-3 to 1e-5 over 50 epochs.

Two gray image test datasets were used to evaluate the performance of the networks. First one is 68 gray images from Berkley segmentation dataset (BSD68), the other one is the collection of 12 gray images commonly used for evaluation of denoising methods. Tables 1,2 show the results of the proposed LFNets compared to state of the art methods on the two test datasets. It can be seen from the results that although there are far fewer learnable parameters in the LFNets, they have achieved better PSNR and SSIM compared to state of the art methods. There are 0.29 million learnable parameters in LFNet_4-5 and 0.35 million learnable parameters in LFNet_4-6 and they outperform DnCNN-S which has 0.56 million learnable parameters. Figure 4 shows the denoising performance of FNets for noise level 50.

LFN_4-5 is able to outperform DnCNN-S [12] with half (52%) the number of learnable parameters.

Figure 3 shows a comparison between the denoising performance of LFNet_4-5 and LFNet_4-6 with two corresponding plain networks which have the same layers as LFNets but layers are stacked one after the other. The conditions of training and testing are the same for the two networks; the only difference is the architecture. The plain networks have 21 and 25 convolutional layers and the patch sizes used for their training are 50 × 50 and 60 × 60, the same as LFNets. To have a more meaningful comparison the number of filters is set to 64 in a few initial layers in plain networks, in order to keep the total number of learnable parameters similar to LFNets, the rest of the layer have 32 filters, similar to LFNets.

It can be seen that even though the deeper 25 layer plain network has more layers, more learnable parameters, and larger receptive field it has failed to improve the results compared to the 21 layer plain network. In contrast the deeper LFNet_4-6 is clearly able to improve the results compared to LFNet_4-5. It also can be seen that the convergence of the LFNets is much faster. After just one epoch LFNet_4-6 achives 26.08 dB, the corresponding

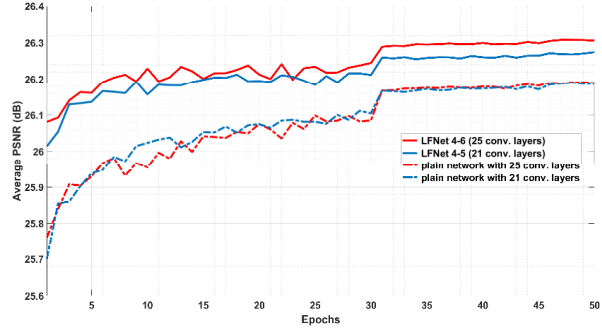

Figure 3. Comparison of the average PSNR (dB) results on BSD68 produced by two LFNets and corresponding two plain networks with the same number of layers as LFNets. Noise level is 50.

plain network reaches this value after 21 epochs.

The ability of LFNets to facilitate the propagation of information and gradient in deep networks for an effective training is clear in Figure 3.

| Noise level | BM3D | WNNM | EPLL | MLP | CSF | TNRD | DnCNN-S (0.56 M param.) | LFNet_4-5 (0.29 M param.) | LFNet_4-6 (0.35 M param.) |
|---|---|---|---|---|---|---|---|---|---|
| σ = 15 | 31.07 | 31.37 | 31.21 | - | 31.24 | 31.42 | 31.72 | 31.73 | 31.76 |
| σ = 25 | 28.57 | 28.83 | 28.68 | 28.96 | 28.74 | 28.92 | 29.23 | 29.24 | 29.26 |
| σ = 50 | 25.62 | 25.87 | 25.67 | 26.03 | - | 25.97 | 26.23 | 26.27 | 26.31 |

Table 1. The average PSNR (dB) of the results of different methods on the BSD68 dataset

| Images | C.man | House | Peppers | Starfish | Monarch | Airplane | Parrot | Lena | Barbara | Boat | Man | Couple | Average |
|---|---|---|---|---|---|---|---|---|---|---|---|---|---|
| Noise Level | | | | | | σ = 15 | | | | | | | |
| BM3D [16] | 31.91 | 34.93 | 32.69 | 31.14 | 31.85 | 31.07 | 31.37 | 34.26 | 33.10 | 32.13 | 31.92 | 32.10 | 32.372 |
| WNNM [17] | 32.17 | 35.13 | 32.99 | 31.82 | 32.71 | 31.39 | 31.62 | 34.27 | 33.60 | 32.27 | 32.11 | 32.17 | 32.696 |
| EPLL [18] | 31.85 | 34.17 | 32.64 | 31.13 | 32.10 | 31.19 | 31.42 | 33.92 | 31.38 | 31.93 | 32.00 | 31.93 | 32.138 |
| CSF [20] | 31.95 | 34.39 | 32.85 | 31.55 | 32.33 | 31.33 | 31.37 | 34.06 | 31.92 | 32.01 | 32.08 | 31.98 | 32.318 |
| TNRD [21] | 32.19 | 34.53 | 33.04 | 31.75 | 32.56 | 31.46 | 31.63 | 34.24 | 32.13 | 32.14 | 32.23 | 32.11 | 32.502 |
| DnCNN-S (0.56 M param.) | 32.61 | 34.97 | 33.30 | 32.20 | 33.09 | 31.70 | 31.83 | 34.62 | 32.64 | 32.42 | 32.46 | 32.47 | 32.859 |
| LFNet_4-5 (0.29 M param.) | 32.64 | 35.01 | 33.24 | 32.19 | 33.08 | 31.71 | 31.85 | 34.60 | 32.62 | 32.41 | 32.47 | 32.48 | 32.859 |
| LFNet_4-6 (0.35 M param.) | 32.65 | 35.03 | 33.28 | 32.25 | 33.13 | 31.71 | 31.87 | 34.65 | 32.72 | 32.45 | 32.49 | 32.51 | 32.895 |
| Noise Level | | | | | | σ = 25 | | | | | | | |
| BM3D [16] | 29.45 | 32.85 | 30.16 | 28.56 | 29.25 | 28.42 | 28.93 | 32.07 | 30.71 | 29.90 | 29.61 | 29.71 | 29.969 |
| WNNM [17] | 29.64 | 33.22 | 30.42 | 29.03 | 29.84 | 28.69 | 29.15 | 32.24 | 31.24 | 30.03 | 29.76 | 29.82 | 30.257 |
| EPLL [18] | 29.26 | 32.17 | 30.17 | 28.51 | 29.39 | 28.61 | 28.95 | 31.73 | 28.61 | 29.74 | 29.66 | 29.53 | 29.692 |
| MLP [19] | 29.61 | 32.56 | 30.30 | 28.82 | 29.61 | 28.82 | 29.25 | 32.25 | 29.54 | 29.97 | 29.88 | 29.73 | 30.027 |
| CSF [20] | 29.48 | 32.39 | 30.32 | 28.80 | 29.62 | 28.72 | 28.90 | 31.79 | 29.03 | 29.76 | 29.71 | 29.53 | 29.837 |
| TNRD [21] | 29.72 | 32.53 | 30.57 | 29.02 | 29.85 | 28.88 | 29.18 | 32.00 | 29.41 | 29.91 | 29.87 | 29.71 | 30.055 |
| DnCNN-S (0.56 M param.) | 30.18 | 33.06 | 30.87 | 29.41 | 30.28 | 29.13 | 29.43 | 32.44 | 30.00 | 30.21 | 30.10 | 30.12 | 30.436 |
| LFNet_4-5 (0.29 M param.) | 30.20 | 33.11 | 30.84 | 29.43 | 30.24 | 29.13 | 29.42 | 32.45 | 29.97 | 30.21 | 30.11 | 30.13 | 30.436 |
| LFNet_4-6 (0.35 M param.) | 30.24 | 33.17 | 30.86 | 29.47 | 30.28 | 29.15 | 29.42 | 32.51 | 30.08 | 30.25 | 30.13 | 30.18 | 30.477 |
| Noise Level | | | | | | σ = 50 | | | | | | | |
| BM3D [16] | 26.13 | 29.69 | 26.68 | 25.04 | 25.82 | 25.10 | 25.90 | 29.05 | 27.22 | 26.78 | 26.81 | 26.46 | 26.722 |
| WNNM [17] | 26.45 | 30.33 | 26.95 | 25.44 | 26.32 | 25.42 | 26.14 | 29.25 | 27.79 | 26.97 | 26.94 | 26.64 | 27.052 |
| EPLL [18] | 26.10 | 29.12 | 26.80 | 25.12 | 25.94 | 25.31 | 25.95 | 28.68 | 24.83 | 26.74 | 26.79 | 26.30 | 26.471 |
| MLP [19] | 26.37 | 29.64 | 26.68 | 25.43 | 26.26 | 25.56 | 26.12 | 29.32 | 25.24 | 27.03 | 27.06 | 26.67 | 26.783 |
| TNRD [21] | 26.62 | 29.48 | 27.10 | 25.42 | 26.31 | 25.59 | 26.16 | 28.93 | 25.70 | 26.94 | 26.98 | 26.50 | 26.812 |
| DnCNN-S (0.56 M param.) | 27.03 | 30.00 | 27.32 | 25.70 | 26.78 | 25.87 | 26.48 | 29.39 | 26.22 | 27.20 | 27.24 | 26.90 | 27.178 |
| LFNet_4-5 (0.29 M param.) | 27.05 | 30.09 | 27.36 | 25.77 | 26.83 | 25.90 | 26.46 | 29.47 | 26.33 | 27.24 | 27.28 | 26.95 | 27.226 |
| LFNet_4-6 (0.35 M param.) | 27.08 | 30.24 | 27.37 | 25.83 | 26.88 | 25.89 | 26.45 | 29.51 | 26.45 | 27.28 | 27.30 | 27.03 | 27.276 |

Table 2. The PSNR (dB) of the results of different methods on 12 commonly used test images

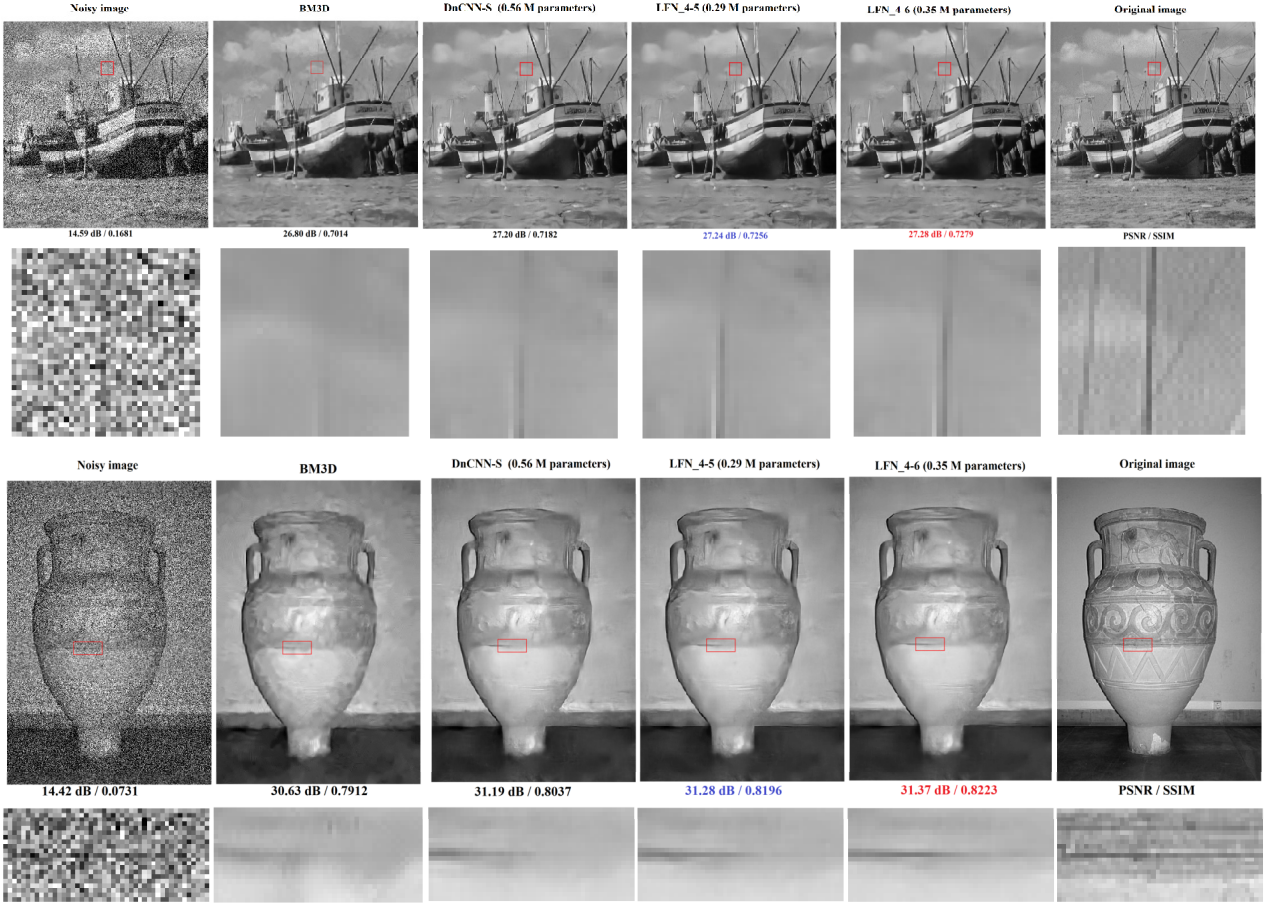

Figure 4. Results of the proposed LFN_4-5 (0.29 M learnable parameters) and LFN_4-6 (0.35 M learnable parameters) compared with Dn-CNN-S (0.56 M learnable parameters). Noise level is 50. Images are from 12 commonly used gray test images.

For color image denoising an LFNet with 4 rows and 10 columns (LFNet_4-10) with 41 convolutional layers in total is implements. This network has 0.61 million learnable parameters. In this network the layers are the same as the gray image denoiser networks, but because the number of image channels is three, in the first layer the size of the filters are $3 \times 3 \times 3$ and last layer has three filters with size $3 \times 3 \times 64$. The receptive field of the network is $83 \times 83$, and the patch size is set to 90.

LFN_4-10 is trained as a blind denoiser for 40 epochs. A single network is used for Gaussian noise levels in the range of [0, 55]. 128×3000 pairs of noisy and residual patches are used as training data. Following [12] training images are 432 color images from Berkley segmentation dataset with 500 color images excluding the 68 color version of BSD68 (CBSD68) which are used as test dataset.

The network is tested using three commonly used test datasets for color denoising. The test datasets are CBSD68, Kodak24 [22], McMaster [23]. Kodak24 and McMaster contain 24 and 18 cropped natural images of size 500×500 respectively. Figures 5,6 and table 3 report the results. The results of LFN_4-10 are compared with CDnCNN-B and CDnCNN-S. Each of these networks has 20 convolutional layers and 0.67 million learnable parameters. CDCNN-B is a blind color denoiser for noise levels in the range of [0, 55] and CDnCNN-S is non-blind and is trained for a specific noise level. CDnCNNs are trained for 50 epochs.

It can be seen from the results that not only LFNet_4-10 (blind) is able to outperform the CDnCNN-B (blind) with a large margin; it is also able to outperform the CDnCNN-S (non-blind) in two of the test datasets. LFNet_4-10, a blind network (trained for 40 epochs), is able to outperform CDnCNN-S, a non-blind network (trained for 50 epochs), with less learnable parameters; this shows the ability of the proposed architecture in effectively training deep neural networks.

| Color Datasets | Methods | σ = 15 | σ = 25 | σ = 35 | σ = 50 |
|---|---|---|---|---|---|
| CBSD68 | CBM3D | 33.52 | 30.71 | 28.89 | 27.38 |
| | CDnCNN-S    (0.67 M param.)  (Non-blind) | 33.98 | 31.31 | 29.65 | 28.01 |
| | CDnCNN-B    (0.67 M param.)  (Blind) | 33.89 | 31.23 | 29.58 | 27.92 |
| | LFNet_4-10_B  (0.61 M param.)  (Blind) | 34.06 | 31.40 | 29.76 | 28.12 |
| Kodak24 | CBM3D | 34.28 | 31.68 | 29.90 | 28.46 |
| | CDnCNN-S    (0.67 M param.)  (Non-blind) | 34.73 | 32.23 | 30.64 | 29.02 |
| | CDnCNN-B    (0.67 M param.)  (Blind) | 34.48 | 32.03 | 30.46 | 28.85 |
| | LFNet_4-10_B  (0.61 M param.)  (Blind) | 34.74 | 32.30 | 30.74 | 29.14 |
| McMaster | CBM3D | 34.06 | 31.66 | 29.92 | 28.51 |
| | CDnCNN-S    (0.67 M param.)  (Non-blind) | 34.80 | 32.47 | 30.91 | 29.21 |
| | CDnCNN-B    (0.67 M param.)  (Blind) | 33.44 | 31.51 | 30.14 | 28.61 |
| | LFNet_4-10_B  (0.61 M param.)  (Blind) | 33.93 | 31.92 | 30.52 | 28.99 |

Table 3. The average PSNR (dB) of the results of different methods on three commonly used color image datasets.

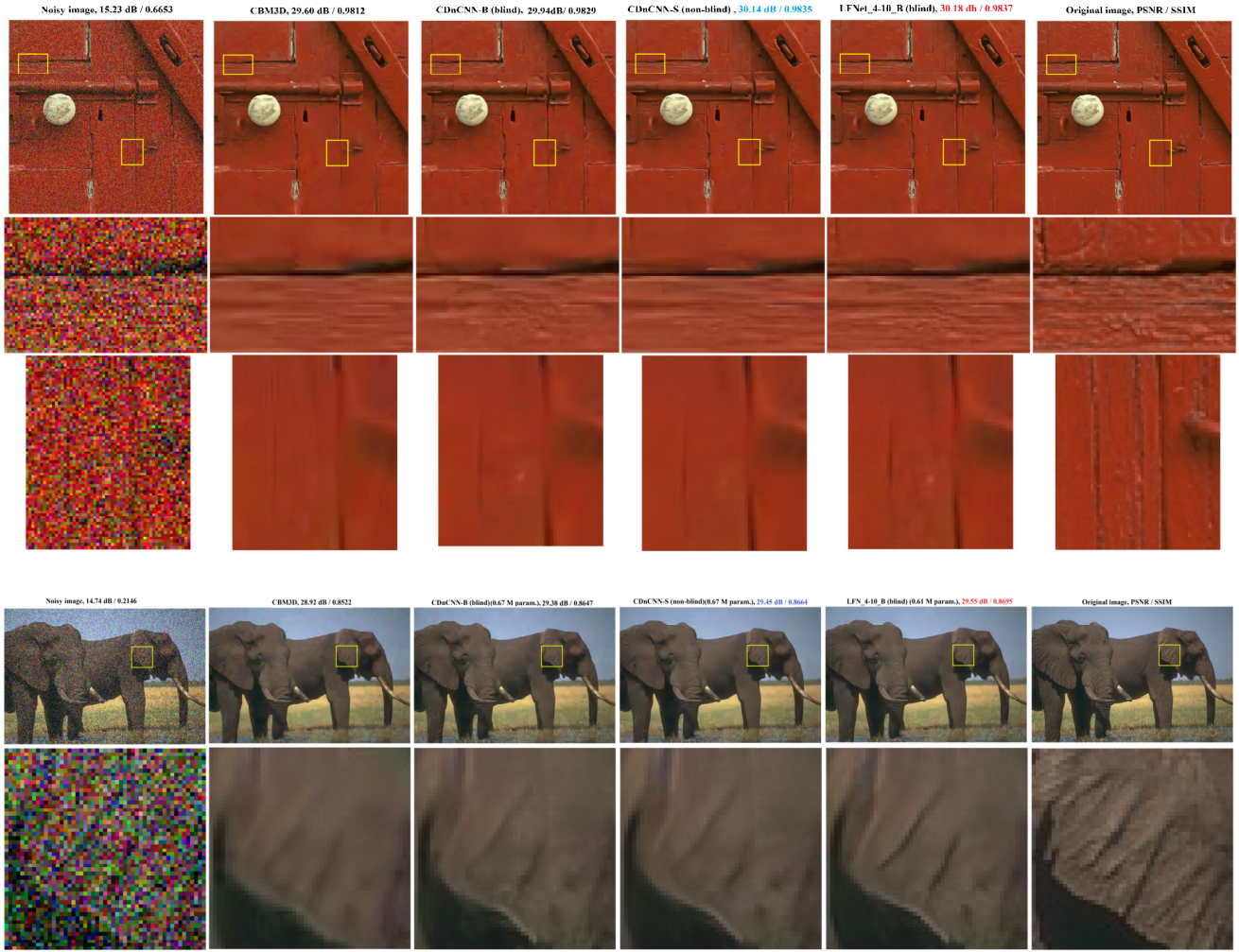

Figure 5. Color image denoising (Gaussian noise) results of different methods. Images are from Kdak24 and CBSD68. Noise level is 50.

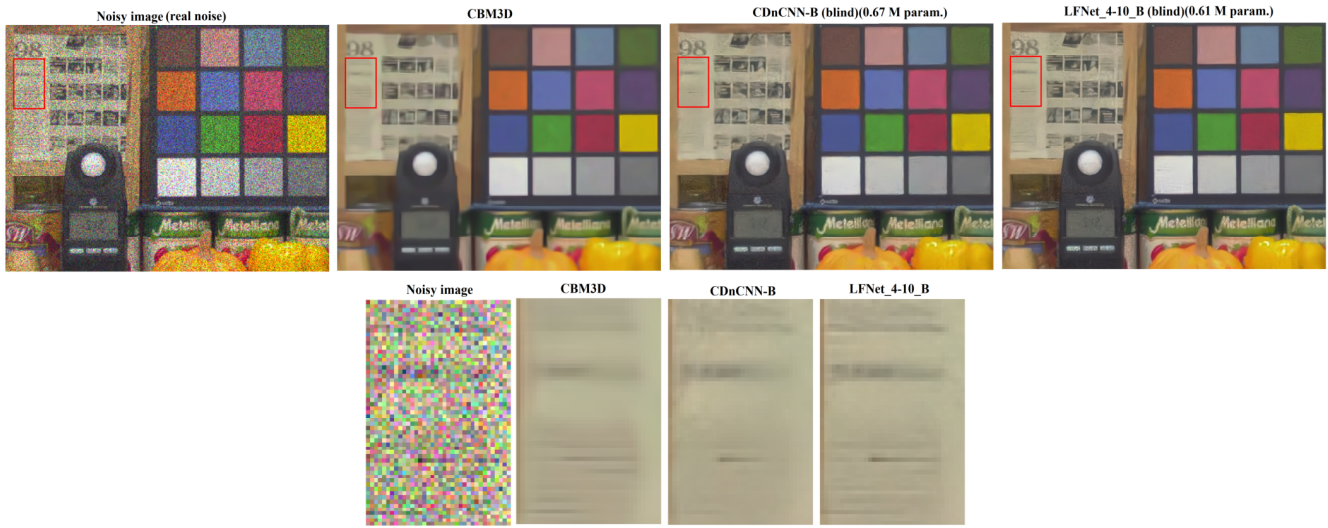

Figure 6. Color image denoising results of different methods. Image is from RNI15 which contains images with real noise.

From the visual comparisons of Figures 4,5,6 for gray and color image denoising it can be seen that LFNets are more successful in recovering image details. This shows the effectiveness of LFNets in training deep networks, and specially reflects the high interaction between feature maps of different levels and the role of multiple depths in LFNets.

## 5. Conclusion and future works

A general framework for deep neural network training is proposed in this paper. In a Lattice Fusion Network (LFNet) convolutional layers are connected based on a directed acyclic lattice graph structures. This structure provides multiple paths, with different depths, for propagation of gradient and information. In LFNet the maximum distance between layers and output is much less than the maximum possible depth of the network, this ensures easier information and gradient propagation even in networks with a large number of layers. Different LFNets were implemented and tested for the task of image denoising to evaluate its performance. LFNets are able to outperform state of the art methods with far fewer number of learnable parameters with shows the effectiveness of the proposed network in training of deep networks.

The main goal of the chosen LFNet architectures here was to prove the effectiveness of LFNets in training deep neural networks, rather than going beyond state of the art methods as much as possible. This area can be further studied in the future.

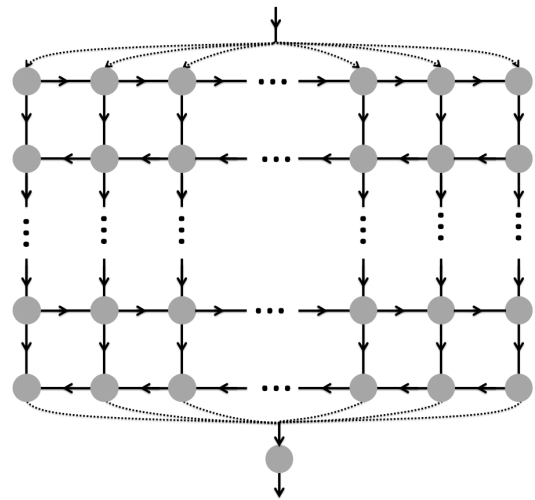

Figure 7. Multiple connections from input and to the output layer to maintain the flow of gradient and information in a LFN with a large number of layers.

A lot of different variants of the general frame work of LFNet can be designed for different machine learning tasks. The example shown in Figure 7 is for a network with very large number of layers. In such a large network the distance of layers to output even with the lattice structure may be too far. To shorten the distance of the layers to input and output layers, multiple layers at the upper and lower part of the network could be connected to the input and output layer to maintain the easy flow of information and gradient.